\documentclass[lettersize,journal]{IEEEtran}
\usepackage{amsmath,amsfonts,color}
\usepackage{algorithmic}
\usepackage{array}
\usepackage[caption=false,font=normalsize,labelfont=sf,textfont=sf]{subfig}
\usepackage{textcomp}
\usepackage{stfloats}
\usepackage{url}
\usepackage{verbatim}
\usepackage{graphicx}
\usepackage{multirow}
\usepackage{cite}
\usepackage{url}
\usepackage{hyperref}
\usepackage{xcolor}
\usepackage{threeparttable}
\def\BibTeX{{\rm B\kern-.05em{\sc i\kern-.025em b}\kern-.08em
    T\kern-.1667em\lower.7ex\hbox{E}\kern-.125emX}}
\usepackage{balance}
\begin{document}
\title{{\scshape VatLM}: Visual-Audio-Text Pre-Training  with Unified Masked Prediction for Speech Representation Learning}
	\author{Qiushi Zhu, Long Zhou, Ziqiang Zhang, Shujie Liu,  Binxing Jiao, Jie Zhang, Lirong Dai, \\ Daxin Jiang, Jinyu Li, Furu Wei
		\thanks{Manuscript created November, 2022; The work was done at Microsoft Research Asia during the internship of the first author. This work is supported by the National Natural Science Foundation of China (62101523), Hefei Municipal Natural Science Foundation (2022012), 
			Fundamental Research Funds for the Central Universities and the Leading Plan of CAS (XDC08010200). (Corresponding author: Long Zhou.)
			
			Qiushi Zhu, Ziqiang Zhang, Jie Zhang and Lirong Dai are with the National Engineering Research Center of Speech and Language Information Processing (NERC-SLIP), University of Science and Technology of China (USTC), Hefei 230026, China (e-mail: qszhu@mail.ustc.edu.cn; zz12375@mail.ustc.edu.cn; jzhang6@ustc.edu.cn; lrdai@ustc.edu.cn).
			
			Long Zhou, Shujie Liu, Binxing Jiao, Daxin Jiang, Jinyu Li and Furu Wei are with the Microsoft Corporation. (e-mail: lozhou@microsoft.com; shujliu@microsoft.com; binxjia@microsoft.com; djiang@microsoft.com; jinyli@microsoft.com; fuwei@microsoft.com)
	}}

\markboth{Journal of \LaTeX\ Class Files,~Vol.~18, No.~9, September~2020}%
{How to Use the IEEEtran \LaTeX \ Templates}

\maketitle
\begin{abstract}

Although speech is a simple and effective way for humans to communicate with the outside world, a more realistic speech interaction contains multimodal information, e.g., vision, text.
How to design a unified framework to integrate different modal information and leverage different resources (e.g., visual-audio pairs, audio-text pairs, unlabeled speech, and unlabeled text) to facilitate speech representation learning was not well explored.
In this paper, we propose a unified cross-modal representation learning framework {\scshape VatLM} (Visual-Audio-Text Language Model).
The proposed {\scshape VatLM} employs a unified backbone network to model the modality-independent information and utilizes three simple modality-dependent modules to preprocess visual, speech, and text inputs.
In order to integrate these three modalities into one shared semantic space, {\scshape VatLM} is optimized with a masked prediction task of unified tokens, given by our proposed unified tokenizer.
We evaluate the pre-trained {\scshape VatLM} on audio-visual related downstream tasks, including audio-visual speech recognition (AVSR), and visual speech recognition (VSR) tasks.
Results show that the proposed {\scshape VatLM} outperforms previous state-of-the-art models, such as the audio-visual pre-trained AV-HuBERT model, and analysis also demonstrates that {\scshape VatLM} is capable of aligning different modalities into the same space.
To facilitate future research, we release the code and pre-trained models at \url{https://aka.ms/vatlm}.

\end{abstract}


\section{Introduction}
\IEEEPARstart{H}{uman} perception of the world is multi-dimensional, including speech, text, pictures, video, touch, taste, etc \cite{865479,6384798,zhu2021deep,9566778}.
Among them, speech is the most natural, simple and effective communication way.
Meanwhile, speech is unique because speech inputs are continuous without a predefined dictionary and vary in the sentence length without segment boundaries \cite{mohamed2022self}.
To alleviate the data scarcity problem of labeled speech, many studies have been dedicated to learning speech representations using self-supervised pre-training methods \cite{Schneider2019,NEURIPS2020_92d1e1eb,9585401,9814838,baevski2022data2vec, oord2018representation,liu2021tera}, which were shown to be beneficial to a variety of spoken language tasks.
Generally speaking, speech representation learning methods can be implemented in contrastive~\cite{oord2018representation,Schneider2019,NEURIPS2020_92d1e1eb}, predictive \cite{9585401,baevski2022data2vec,9814838} and generative fashions \cite{Chung2019,liu2021tera,xu2022masked}.
For example, wav2vec2.0~\cite{NEURIPS2020_92d1e1eb} utilizes a bidirectional Transformer~\cite{vaswani2017attention} structure and performs a contrastive learning task over a 
quantization of speech representations.
HuBERT \cite{9585401} iteratively clusters MFCC features or the representation output from the middle layer of the pre-trained model to generate targets for self-supervised pre-training, like BERT pre-training \cite{9688253}.
Audio-MAE \cite{xu2022masked} extends image-based masked auto-encoder (MAE) \cite{he2022masked} to self-supervised speech representation learning by reconstructing audio spectrograms.

Recent years have also witnessed a remarkable progress in the enhanced speech representation learning by using additional visual or textual information \cite{Xu_2020_CVPR,9004036,bapna2022mslam,zhang2022speechlm}.
On the one hand, speech production is accompanied by lip movements, where both audio and visual information facilitate speech understanding process of human \cite{sumby1954visual}.
Additional visual information has shown to be able to improve the automatic speech recognition (ASR) performance, especially in noisy environments~\cite{8461326,8585066,9414567,Xu_2020_CVPR,9004036,9018185,4540195}.
Moreover, visual information can facilitate communication for people with speech disorders.
On the other hand, text and speech have a natural alignment relationship \cite{zhang2022speechlm}, although they are two different modalities. Joint pre-training of speech and text is therefore beneficial for speech representation learning as well as subsequent cross-modal tasks, such as speech recognition, speech translation, and speech synthesis \cite{bapna2021slam,bapna2022mslam,ao-etal-2022-speecht5,zhang2022speechlm}.
Based on the previous discussion, one question should be asked: \textit{how to effectively integrate both visual information and textual information into the single speech representation learning model?}

To answer this question, multimodal speech pre-training is explored in these years and is usually categorized into audio-visual representation learning and audio-text representation learning.
For the audio-visual representation learning,  the complementary information contained in the video modality can be exploited to enhance the speech representation~\cite{shi2022learning,hsu2022single,pan-etal-2022-leveraging,zhang2022learning,shi2022robust}.
Based on a self-supervised pre-training method, AV-HuBERT~\cite{shi2022learning}  learns the representations of two modalities by masked prediction of automatically discovered and iteratively refined multimodal hidden units.
The authors in \cite{shi2022robust} aim at improving the robustness of AV-HuBERT with noise-augmented pre-training and fine-tuning approaches.
In these audio-visual representation learning methods, the text information was not considered, despite the text data are abundant and might be beneficial for audio-visual learning tasks.
For the audio-text speech representation learning, the goal is to boost the quality of the speech representation with additional unlabeled text data \cite{bapna2021slam,ao-etal-2022-speecht5,tang-etal-2022-unified,zhang2022speechlm, weiwang2021}.
For example, in SLAM~\cite{bapna2021slam} a single encoder is constructed in a multi-task learning framework with the BERT objective for unlabeled text together with the w2v-BERT~\cite{9688253} objective for unlabeled speech.
Besides, SpeechLM~\cite{zhang2022speechlm} aligns unlabeled speech and text pre-training in a pre-defined discrete representation with a shared encoder network.
Similarly, previous audio-text speech representation learning methods can only leverage speech and text data without considering visual information. 
With the visual or text information, previous multimodal speech pre-training enhances the speech representation and improves the performance of downstream tasks. However, they exhibit the following disadvantages.
\begin{itemize}
\item Previous multimodal (visual, speech, text) pre-training approaches \cite{wang2022image,liu2021opt,yang2022code} mainly focus on visual-language tasks and have not been validated on some spoken language tasks, such as AVSR.
\item Previous speech representation learning methods can not make full use of diverse corpora, e.g., visual-audio pairs, audio-text pairs, and unlabeled speech and text, without considering both the visual and textual information.
\item Previous methods mostly depend on a complicated model architecture and pre-training objects, lacking a unified multimodal framework for different modalities  modeled in the same semantic space.
\end{itemize}

To address these problems, 
we propose a unified pre-training framework, Visual-Audio-Text Language Model ({\scshape VatLM}), to model the representations of three modalities using a unified architecture and masked prediction objectives. 
More specifically, we first design a multipath Transformer consisting of three pre-processing modules for visual, audio and text, respectively, and a shared Transformer model to learn modality-independent representation.
Second, we propose a unified tokenizer to convert the corpora from different modalities to hidden units in the same semantic space. Following AV-HuBERT \cite{shi2022learning}, visual and audio samples are discretized with the audio-visual clustering model of the fourth iteration AV-HuBERT. Inspired by SpeechLM \cite{zhang2022speechlm}, we introduce a text-to-unit model trained on a small speech-transcription dataset to translate unpaired text data to discrete units.
Third, with the converted hidden units as targets, we adopt a simple but effective masked prediction pre-training task, like BERT \cite{9688253}, for unimodal and multimodal data (unlabeled audio and unlabeled text, paired visual-speech, and paired speech-text). 

Our proposed {\scshape VatLM} is evaluated on AVSR and VSR tasks, and experimental results show the superiority and effectiveness of our proposed {\scshape VatLM} compared with existing methods, e.g., the state-of-the-art AV-HuBERT model. Particularly, with only 30h labeled LRS3 data, {\scshape VatLM} Base obtains the 10.5\% and 8.4\% word error ratio (WER) reduction than AV-HuBERT in AVSR (3.4 vs. 3.8) and VSR (42.6 vs. 46.1) tasks, and {\scshape VatLM} Large achieves the performance of 2.7 and 31.6 WER for AVSR and VSR, respectively.
Following the experiment results, ablation study, robustness evaluation, and visualization analysis are conducted for deep analyses.

The contributions of this paper are summarized as follows,
\begin{itemize}
\item We propose a unified visual-audio-text speech representation model ({\scshape VatLM}), which employs a shared Transformer network and a masked prediction task to learn a unified representation of visual, audio, and text.
\item We empirically evaluate the proposed {\scshape VatLM} on different downstream tasks, including audio-visual speech recognition and visual speech recognition tasks. Experimental results and analysis demonstrate the effectiveness and superiority of our {\scshape VatLM}.
\item The proposed {\scshape VatLM} is simple but very effective, and this framework is easy to extend to more corpora from different modalities and larger model capacity with more parameters. Code and models are available at \url{https://aka.ms/vatlm}.
\end{itemize}

The remainder of the paper is organized as follows. We discuss the related work in Section \ref{sec:RelatedWork}. We describe the proposed method in Section \ref{sec:method} in detail. Section \ref{sec:experiment_setting} and Section \ref{sec:experiment_results} show the experimental setup and results, respectively.  Finally, we give a brief conclusion in Section \ref{sec:conclusion}.

\section{Related Work}
\label{sec:RelatedWork}


Our work is based on self-supervised speech representation learning, and it is also related to cross-modal representation learning with visual, audio, and text. We discuss these topics in the following.

\subsection{Speech Representation Learning}
Self-supervised pre-training approaches have made remarkable success in speech representation learning in recent years.
Based on the training objectives, these approaches can be categorized into contrastive methods,  predictive methods, and generative methods \cite{Chung2019,oord2018representation,Schneider2019,9585401,baevski2022data2vec,9814838,xu2022masked}.
Autoregressive predictive coding (APC)~\cite{Chung2019} was proposed to reconstruct future frames based on past frames. Contrastive predictive coding (CPC)~\cite{oord2018representation} and wav2vec~\cite{Schneider2019} perform the next-step prediction in a similar way but using a contrastive loss. 
wav2vec2.0~\cite{NEURIPS2020_92d1e1eb} utilizes a bidirectional Transformer~\cite{vaswani2017attention} structure, and performs mask prediction task via a contrastive task on the quantized representation.
In wav2vec2.0~\cite{NEURIPS2020_92d1e1eb}, local features are employed as targets for self-supervised pre-training, where the contextual information is not fully leveraged. This problem was then considered in HuBERT~\cite{9585401} and data2vec~\cite{baevski2022data2vec}.
Like BERT pre-training, HuBERT is optimized with self-supervised masked prediction task, where the target labels come from offline clusters of the representations of the previous pre-trained model or MFCC features.
On the basis of HuBERT, WavLM~\cite{9814838}
utilizes a sentence-level mixing data augmentation approach to enhance the speaker information, which performs very well on the
SUPERB benchmark~\cite{yang21c_interspeech}.
Inspired by image-based Masked Autoencoders (MAE) \cite{he2022masked}, Audio-MAE \cite{xu2022masked} reconstructs audio spectrograms to learn self-supervised speech representation. 

\begin{figure*}[!t]
	\centering
	\includegraphics[width=0.9\textwidth]{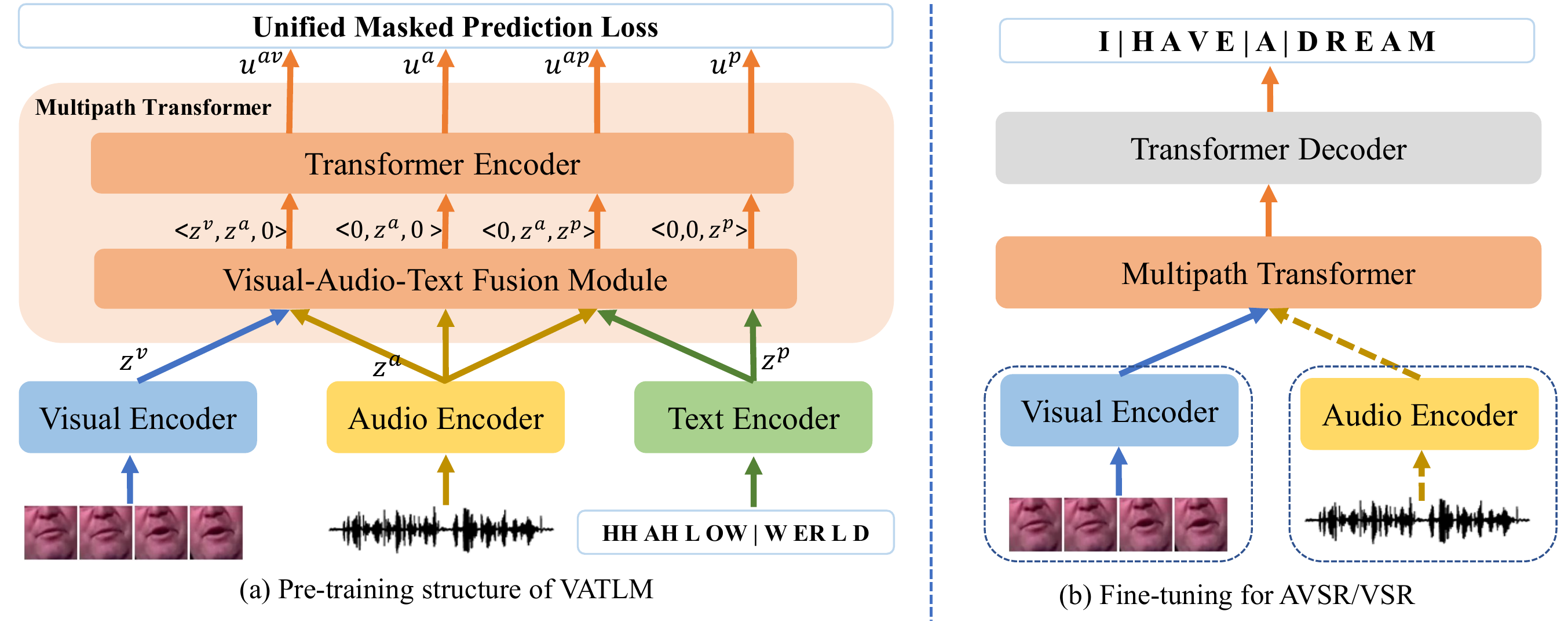}
	\caption{An illustration of the {\scshape VatLM} model: (a) pre-training structure of {\scshape VatLM} and (b) fine-tuning structure for AVSR/VSR tasks.}
	\label{fig:vatlm}
\end{figure*}

\subsection{Cross-Modal Learning}

\textbf{Visual-Audio Learning}
Visual information is complementary to the audio signal in many scenarios, so visual-audio speech representation learning has obtained more and more attention, especially for speech recognition tasks \cite{9054253,ma21c_interspeech,pan-etal-2022-leveraging,zhang2022learning,9746449,Prajwal_2022_CVPR,kim2022distinguishing,ma2022visual}.
In~\cite{9054253}, A VSR model is trained on a large amount of unlabeled data, which is achieved by distilling the trained ASR model.
LiRA~\cite{ma21c_interspeech} leverages audio modality as a training objective to learn visual speech representations via a self-supervised approach.
AV-BERT~\cite{9746449} focuses on learning utterance-level multimodal environment embedding that serves as the global context for ASR.
In~\cite{Prajwal_2022_CVPR}, an attention-based pooling mechanism was proposed to aggregate visual speech representations and use sub-word units for lip reading.
In~\cite{kim2022distinguishing}, Multi-head Visual-audio Memory (MVM) was proposed, which can extract possible candidate audio representations from one visual lip movement to mitigate the effect of homophones.
The authors in ~\cite{ma2022visual} propose to add prediction-based auxiliary tasks to the VSR model and emphasize the importance of hyperparameter optimization and data augmentation.
Both AV-HuBERT \cite{shi2022learning} and u-HuBERT \cite{hsu2022single} are self-supervised audio-visual representation learning frameworks, which predict target cluster labels consuming masked visual and audio frames, and achieve state-of-the-art performance on AVSR and VSR tasks.
u-HuBERT~\cite{hsu2022single} utilizes noise augmentation methods with additional unimodal audio data in the pre-training phase, and pre-trains the model with more updates (1000k). 
In \cite{shi2022robust}, the authors further improve the robustness of audio-visual speech recognition against different types of noises.

\textbf{Audio-Text Learning}
Although there is a large difference between speech modality and text modality, speech and text are two expressions of language. Hence, it is a promising direction to investigate how to bridge the modality gap between speech and text \cite{bapna2021slam,chung-etal-2021-splat,ao-etal-2022-speecht5,tang-etal-2022-unified,zhang2022speechlm}.
The authors in \cite{ao-etal-2022-speecht5} propose a unified-modal encoder-decoder pre-training framework SpeechT5, which converts most spoken language tasks into speech/text to speech/text tasks, and utilizes large-scale unlabeled speech and text data to pre-train the model.
Based on the RNN-T framework \cite{zhang2020transformer}, MAESTRO \cite{chen2022maestro} model tries to learn a unified representation of speech and text through a modality matching algorithm.
SLAM~\cite{bapna2021slam} and mSLAM \cite{bapna2022mslam} are proposed to model speech and text with a single encoder, which utilizes the paired speech-transcription corpus to learn the alignment relationship.
In \cite{zhang2022speechlm}, the authors propose a text-augmented speech pre-trained model SpeechLM, which utilizes phoneme units or hidden units as the bridge of speech and text modalities to achieve excellent performance.

\textbf{Visual-Audio-Text Learning}
There is a big convergence in model architectures and training objects for different modalities, e.g., visual, audio, and text \cite{wang2022image,liu2021opt,yang2022code}. 
The authors in \cite{akbari2021vatt} propose a VATT model, which takes original signals of different modalities as inputs and employs multi-modal contrastive learning to optimize the model. The proposed VATT model outperforms ConvNet-based architectures in video action recognition, audio event classification, image classification, and text-to-video retrieval tasks.
BEIT-3 \cite{wang2022image} is a general-purpose multimodal foundation model, which models images as a foreign language, and employs a masked language modeling task for unpaired images, unpaired texts, and paired image-text.
To address the challenge that the algorithms and objects usually differ widely in different modalities, a general self-supervised learning framework data2vec \cite{baevski2022data2vec} is proposed to separately predict the latent representation of original input by the base model based on a masked view of the input.

Different from existing work, the goal in this paper is to design a universal multimodal (visual, speech, and text) model for speech representation learning, with a unified pre-training object to leverage different data sources, including  paired visual-audio, audio-text, and unpaired audio and text\footnote{We do not consider unpaired visual corpus because visual data usually contains audio data, and they forms paired visual-audio data.}.

\section{VatLM}
\label{sec:method}

In this section, we will introduce the overall framework of {\scshape VatLM} and how to pre-train {\scshape VatLM} in the same semantic space.
As shown in Fig.~\ref{fig:vatlm}, the {\scshape VatLM} model contains a visual preprocessing module, an audio preprocessing module, a text preprocessing module (Section \ref{sec_Preprocessing_Module}) and a shared multipath Transformer module (Section \ref{sec_Backbone_Network}).
Using a unified tokenizer (Section \ref{sec_Discretization_Method}), we can pre-train {\scshape VatLM} as a unified masked prediction task (\ref{sec_Pre_Training_Task}) for different modalities.
Followed with fine-tuning, the pre-trained {\scshape VatLM} can be adapted to various downstream tasks, such as AVSR and VSR.


\subsection{Preprocessing Module: Visual/Audio/Text Encoder}
\label{sec_Preprocessing_Module}
Visual, audio and text modalities have very different characteristics. For example, visual is a three-dimensional (3D) continuous signal, but both audio and text are 1D continuous and discrete signals. Besides, unlike textual data, speech and visual inputs do not have segment boundaries and predefined dictionaries. Hence, following \cite{shi2022learning, ao-etal-2022-speecht5}, we introduce three modality-independent pre-processing modules, i.e., visual/audio/text encoders.

More specifically, for the image frames in the video data, we use dlib~\cite{king2009dlib} to detect 68 facial keypoints and extract a 96$\times$96 region-of-interest (ROI) centered by the mouth, which is then fed to the visual encoder to compute the visual features $z^v$.
Similarly to~\cite{martinez2020lipreading,9414567,shi2022learning}, we use a modified ResNet-18 as the visual encoder, where the first convolution layer is replaced by a 3D convolution with a kernel size of 5$\times$7$\times$7.
Finally, we flatten the visual features corresponding to each video frame into a 1D vector by a two-dimensional average pooling layer.
For the audio data, we extract the 26-dimensional log-filterbank features from the original waveform.
Since the lip frames sampled at 25 Hz and the filterbank feature frames at 100 Hz, we concatenate four adjacent filterbank frames to synchronize the audio and video modalities.
The concatenated filterbank frames are fed to the audio encoder to get the audio features $z^a$.
To avoid the over-reliance issue on the audio stream, we use a single linear project layer as in~\cite{shi2022learning}  to encode the audio input, forcing the audio encoder to learn simple features.
Moreover, we integrate textual data into the pre-trained model by introducing a text encoder module, implemented with an embedding layer.
To ensure that the text feature granularity is approximately the same as the audio feature granularity, we first convert the text sequence into a phoneme sequence, which is then fed to the text encoder to generate the phoneme features $z^p$.



\begin{figure*}[!t]
	\centering
	\includegraphics[width=0.9\textwidth]{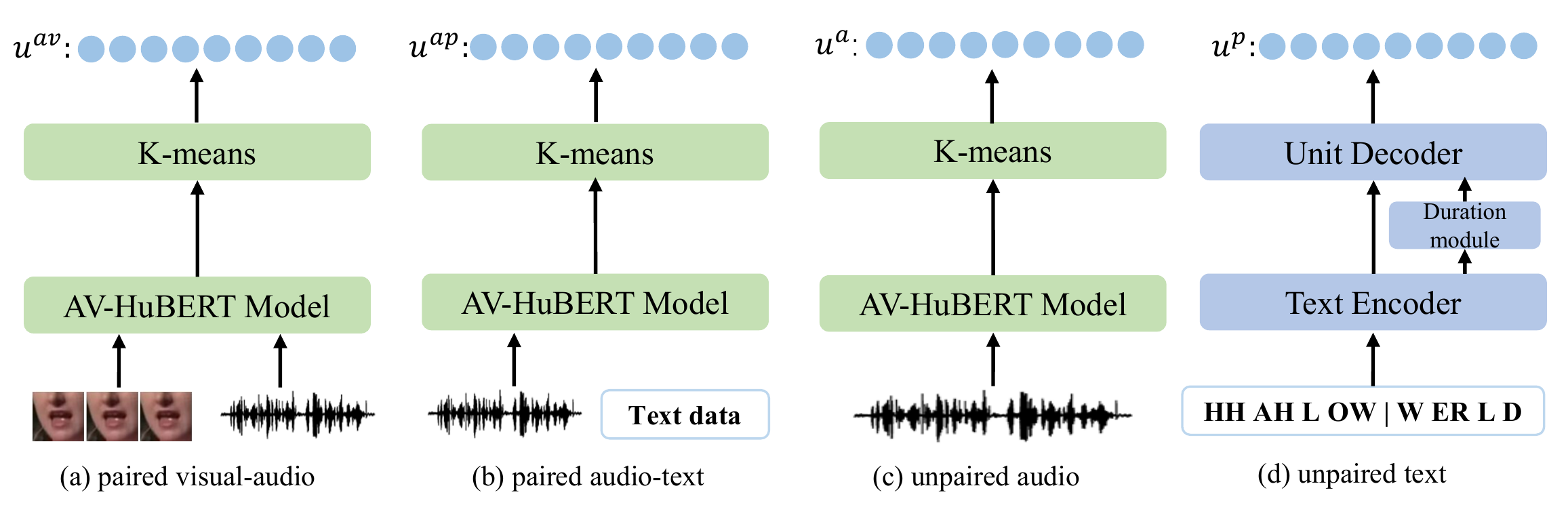}
	\caption{ The unified tokenizer that generates the shared hidden units from different modalities and data resources, including (a) paired visual-audio, (b) paired audio-text, (c) unpaired audio, and (d) unpaired text.}
	\label{fig:avtokenizer}
\end{figure*}

\subsection{Backbone Network: Multipath Transformer}
\label{sec_Backbone_Network}
The big convergence of model architecture across vision, audio, text, and multimodal pre-training has become a mainstream trend \cite{ao-etal-2022-speecht5,shi2022learning,wang2022image}.
In this work, we use an improved multipath Transformer from \cite{shi2022learning} as the basic backbone model to encode different modalities, including visual, audio and text.
Multipath Transformer contains a visual-audio-text fusion module and a Transformer encoder, as shown in Fig. \ref{fig:vatlm}(a). Each Transformer encoder block consists of a self-attention network and a feed-forward network, which are boosted with positional embedding and layer normalization \cite{vaswani2017attention}.

In order to better integrate different modalities in pre-training and be applied to various downstream tasks, we utilize an audio-visual-text fusion module, which simply concatenates the vector representations of visual, audio and text. 
Visual features $z^v$, audio features $z^a$ and text features $z^p$ are concatenated to obtain $z^f$, i.e.,
$z^f = {\rm concat}(z^v, z^a, z^p)$.
For the multimodal features, in case the data of one modality is not available, we replace the corresponding features with zero vectors.
In addition, we also adapt the modality dropout strategy as in~\cite{shi2022learning}, where the input of a modality is randomly dropped during training and the input of the dropped modality is replaced with a zero vector.
The masked feature $z^f$ is sent to the Transformer encoder to learn the contextual feature $h^f = {\rm{Transformer}}(z^f)$.

\subsection{Discretization Method: Unified Tokenizer}
\label{sec_Discretization_Method}

To align visual, audio, and text into the same semantic space, we propose a unified tokenizer to tokenize three-modality data into the shared discrete hidden units.
These multimodal discrete units serve as targets in the pre-training stage.
Specifically, we generate a unified unit for video-audio data, audio-text data, unimodal audio data and unimodal text data.
First, for video-audio data, the features are obtained by feeding the video-audio data into the AV-HuBERT model pre-trained in the fourth iteration\footnote{The reason for using the fourth iteration of AV-HuBERT is that the AV-HuBERT pre-training was done for a total of five iterations, while we trained our {\scshape VatLM} model only once to save the pre-training time.} and generating the hidden units $u^{av}$ by k-means clustering~\cite{9585401}, as shown in Fig.~\ref{fig:avtokenizer}(a), consistent with~\cite{shi2022learning}.
For ease of presentation, we denote the input data and the target label data in brackets ($\langle$video input, audio input, text input$\rangle$, target), and we obtain pairs of data ($\langle$$z^{v}, z^{a}$, 0$\rangle$, $u^{av}$), where 0 means zero vectors in case the text input is not available.
Second, for unimodal audio data, we replace the video modal data with zero vectors, feed the audio data into the pre-trained AV-HuBERT model to get the features, and use the shared k-means model to generate the features into unit $u^a$, such that the paired data ($\langle$0, $z^a$, 0$\rangle$, $u^a$) can be constructed as shown in the Fig.~\ref{fig:avtokenizer}(c).
Third, for paired audio-text data (see Fig. \ref{fig:avtokenizer}(b)), we similarly use the unimodal audio data to obtain unit $u^{ap}$ regardless of the text in extracting target labels, where the corresponding text data is converted into a sequence of phonemes, such that data pairs ($\langle$0, $z^a$, $z^p$$\rangle$, $u^{ap}$) are formed.

Finally, for unpaired text data, we employ a sequence-to-sequence phoneme2unit model to unify the text data to the same hidden-unit space, like \cite{zhang2022speechut,zhang2022speechlm}.
Concretely, following FastSpeech~\cite{NEURIPS2019_f63f65b5}, the phoneme2unit model contains a text encoder, a duration module and a unit decoder, as shown in Fig.~\ref{fig:avtokenizer}(d).
We first train the phoneme2unit model using a small amount of paired phoneme-unit data from the paired ASR speech-transcription data, where the ASR transcription sequence is converted into a phoneme sequence by looking up the lexicon, and the ASR speech data is converted to hidden units using a tokenizer as in Fig~\ref{fig:avtokenizer} (c).
Based on the pre-trained phoneme2unit model, the large-scale unpaired text data are transformed into phoneme sequence, and are fed into the phoneme2unit model to generate the corresponding units. As such, the paired data ($\langle$0, 0, $z^p$$\rangle$, $u^p$) can be obtained.
For the different-modality data, the corresponding hidden units are obtained by using a unified tokenizer with shared semantic labels, which will facilitate our unified masked prediction task.
We demonstrate different modalities indeed share the common representation space in Section \ref{sec:VisualizationAnalysis}.


\subsection{Pre-Training Task: Unified Masked Prediction}
\label{sec_Pre_Training_Task}

Existing multimodal pre-training work usually employs multiple pre-training objectives, such as masked language modeling, contrastive learning, speech-text matching, image-text matching, and so on.
In contrast, we pre-train {\scshape VatLM} via a unified masked prediction objective on both mono-modal (i.e., audio and texts) and multimodal data (i.e., audio-visual pairs and audio-text pairs). The masked prediction objective uses a unit-based masking language modeling task similarly to HuBERT~\cite{9585401}.
The masked multimodal feature $z^f$ is fed into the Transformer encoder to learn the contextual representation $h^f$.
For the contextual representation $h^f=(h_1^f, ..., h_M^f)$, we aim to predict the hidden unit $u={(u_1, ..., u_M)}$ of the corresponding modality at the masked position.
The probability of the predicted unit can be calculated by 
\begin{equation}
	p\left(u \mid h_{t}^{f}\right)=\frac{\exp \left(\text{sim} \left(W h_{t}^{f}, \boldsymbol{e}(u)\right) / \tau\right)}{\sum_{u^{\prime} \in \mathcal{U}} \exp \left(\text{sim} \left(W h_{t}^{f}, \boldsymbol{e}\left(u^{\prime}\right)\right) / \tau\right)},
\end{equation}
where sim($\cdot$) denotes the cosine similarity between two vectors, $W$ is a projection matrix, $\mathbf{e}$($\cdot$) is an embedding matrix, $\tau$ is a temperature coefficient, and $\mathcal{U}$ is the set of hidden units.
The {\scshape VatLM} pre-training loss can be formulated as
\begin{equation}
\mathcal{L} = -\sum_{t\in \mathcal{M}} \left(\log p(u_t|h_t^f)\right),
\end{equation}
where $u_t$ denotes the hidden unit corresponding to the modality at position $t$ and $\mathcal{M}$ is the set of masked positions.
The overall loss function can be expressed as \begin{equation}
	\mathcal{L}_{total} = \mathcal{L}^{av} +\lambda_1\mathcal{L}^a + \lambda_2\mathcal{L}^{ap} + \lambda_3\mathcal{L}^p ,
	\label{eq3}
\end{equation}
where $\mathcal{L}^{av}$ is the loss function for the video-audio data, $\mathcal{L}^{a}$ for the unimodal audio data, $\mathcal{L}^{ap}$ for the paired audio-text data, and $\mathcal{L}^{p}$ for the unimodal text data, respectively, and $\lambda_1$, $\lambda_2$, $\lambda_3$ are the weight hyperparameters. 



\section{Experimental Setting}
\label{sec:experiment_setting}
\subsection{Data Description}

\textbf{Pre-training data}:
	The LRS3\footnote{\url{https://www.robots.ox.ac.uk/~vgg/data/lip_reading/lrs3.html}}~\cite{afouras2018lrs3} and Voxceleb2\footnote{\url{https://www.robots.ox.ac.uk/~vgg/data/voxceleb/vox2.html}}~\cite{chung2018voxceleb2} datasets are utilized as our video-audio data.
	LRS3 is a publicly available sentence-level lip-reading dataset containing 433 hours of videos extracted from TED English talks.
	The original dataset contains two parts: pretrain (403 hours) and trainval (30 hours), both of which are transcribed at the sentence level and have the same source as the test set.
	In the pre-training phase, a total of 433 hours of data is used as unlabeled data.
    The Voxceleb2 dataset is originally created for multilingual audio-visual speaker recognition, containing 2442-hour data in total and extracted from YouTube videos without ground-truth transcriptions.
	The Voxceleb2 dataset has significant domain differences from LRS3, because it is multilingual and contains interviews, talks, etc., in both indoor and outdoor environments.
	Therefore, we only use the 1326-hour English part of Voxceleb2 for pre-training, where the English data selected from Voxceleb2 in the same way as in~\cite{shi2022learning}.
	For the unpaired audio data, we use the GigaSpeech\footnote{\url{https://github.com/SpeechColab/GigaSpeech}}~\cite{chen2021gigaspeech} dataset containing multi-domain audios.
	GigaSpeech contains a total of 10,000 hours of transcribed English data with 5 training subsets, namely XS, S, M, L, and XL.
	Since most of the LRS3 and Voxceleb2 data are originated from YouTube, we only use 3846 hours of YouTube unlabeled audio data from the GigaSpeech training subset XL.
	For the paired audio-text data, we use the TED-LIUM3\footnote{\url{https://lium.univ-lemans.fr/ted-lium3/}}~\cite{tedlium3} dataset, which contains 452 hours of talk audios (extracted from the TED website) in English with ground-truth transcriptions.
	For the unlabeled text data, we use the English text dataset Cantab-TEDLIUM Release 1.1\footnote{\url{https://www.openslr.org/27/}}~\cite{7179001} published by Cantab Research for training neural language models, which contains a total of about 600M English sentences.

\textbf{Fine-tuning data}:
	For the AVSR/VSR downstream task, to validate the performance of the pre-trained model in a low-resource scenario, we fine-tune the pre-trained model with 30 hours of LRS3 labeled subset data.
	In addition, we utilize 433 hours of LRS3 labeled data for fine-tuning in order to compare with other methods.
	To verify the robustness of the model in noisy scenarios, we also fine-tune it on noisy datasets to test the performance in different tasks.
	As in~\cite{shi22_interspeech}, we adopt the noise dataset MUSAN\footnote{\url{http://www.openslr.org/17/}}~\cite{snyder2015musan}, which contains a total of 109 hours of speech, music and babble noise.
	Besides, we also validate our method on the LRS2\footnote{\url{https://www.robots.ox.ac.uk/~vgg/data/lip_reading/lrs2.html}}~\cite{8585066} dataset, which contains 224 hours of audio-visual transcribed data extracted from BBC television programs.

\begin{table*}[]
	\caption{ Comparison of the AVSR/VSR  performance in WER with previous works on the LRS3 dataset.}
	\label{tab:table1}
	\centering
	\begin{threeparttable}
	\begin{tabular}{lccccccc}
		\hline
		\textbf{Method}                               & \textbf{Backbone}                                      & \textbf{Criterion}  & \textbf{\begin{tabular}[c]{@{}c@{}}Extra\\ labeled (hrs)\end{tabular}} & \textbf{\begin{tabular}[c]{@{}c@{}}Fine-tuned\\ data (hrs)\end{tabular}} & \textbf{\begin{tabular}[c]{@{}c@{}}Pre-trained\\ AV data (hrs)\end{tabular}} & \textbf{\begin{tabular}[c]{@{}c@{}}AVSR\\ WER (\%)\end{tabular}} & \textbf{\begin{tabular}[c]{@{}c@{}}VSR\\ WER (\%)\end{tabular}} \\ \hline
		\textbf{Supervised}                           &                                                        &                     &                                                                        &                                                                          &                                                                              &                                                                  &                                                                 \\
		Zhang et al.~\cite{Zhang_2019_ICCV}                                  & CNN                                                    & CE                  & 157                                                                    & 698                                                                      & -                                                                            & -                                                                & 60.1                                                            \\
		Afouras et al.~\cite{8585066}                                & Transformer                                            & CE                  & 157                                                                    & 1362                                                                     & -                                                                            & 7.2                                                              & 58.9                                                            \\
		Xu et al.~\cite{Xu_2020_CVPR}                                     & RNN                                                    & CE                  & 157                                                                    & 433                                                                      & -                                                                            & 7.2                                                              & 57.8                                                            \\
		Shillingford et al.~\cite{shillingford19_interspeech}                            & RNN                                                    & CTC                 & -                                                                      & 3886                                                                     & -                                                                            & -                                                                & 55.1                                                            \\
		Ma et al.~\cite{9414567}                                     & Conformer                                              & CTC+CE              & -                                                                      & 433                                                                      & -                                                                            & -                                                                & 46.9                                                            \\
		Ma et al.~\cite{9414567}                                     & Conformer                                              & CTC+CE              & 157                                                                    & 433                                                                      & -                                                                            & 2.3                                                              & 43.3                                                            \\
		Prajwal et al.~\cite{Prajwal_2022_CVPR} & Transformer & CE & - & 698 & - & - & 40.6
        \\
        Prajwal et al.~\cite{Prajwal_2022_CVPR} & Transformer & CE & - & 2676 & - & - & 30.7
		\\
        Ma et al.~\cite{ma2022visual} & Transformer & CTC+CE & - & 433 & - & - & 37.9
		\\
        Ma et al.~\cite{ma2022visual} & Transformer & CTC+CE & - & 1459 & - & - & 31.5
        \\
		Makino et al.~\cite{9004036}                                 & RNN                                                    & Transducer          & -                                                                      & 31000                                                                    & -                                                                            & -                                                                 & 33.6                                                            \\ \hline
		\textbf{Self-supervised \& Semi-supervised}   &                                                        &                     &                                                                        &                                                                          &                                                                              &                                                                  &                                                                 \\
		Afouras et al.~\cite{9054253}                                & CNN                                                    & CTC                 & 157                                                                    & 433                                                                      & 334                                                                          &                                                                  & 59.8                                                            \\ \hline
		Zhang et al.~\cite{zhang2022learning}                                  & \multicolumn{1}{l}{Transformer-Base}                   & CTC                 & -                                                                      & 30                                                                       & 433                                                                          & 9.1                                                              & 67.8                                                            \\ \hline
		\multirow{2}{*}{Ma et al.~\cite{ma21c_interspeech}}                    & \multirow{2}{*}{Transformer-Base}                      & \multirow{2}{*}{CE} & -                                                                      & 30                                                                       & 433                                                                          & -                                                                & 71.9                                                            \\
		&                                                        &                     & -                                                                      & 433                                                                      & 1759                                                                         & -                                                                & 49.6                                                            \\ \hline
		\multirow{6}{*}{AV-HuBERT~\cite{shi2022learning,shi2022robust}}                    & \multirow{3}{*}{Transformer-Base}                      & \multirow{3}{*}{CE} & -                                                                      & 30                                                                       & 433                                                                          & -                                                              & 51.8                                                            \\
		&                                                        &                     & -                                                                      & 30                                                                       & 1759                                                                         & 4.0                                                              & 46.1                                                            \\
		&                                                        &                     & -                                                                      & 433                                                                      & 1759                                                                         & -                                                              & 34.8                                                            \\ \cline{2-8} 
		&      \multirow{2}{*}{Transformer-Large}                                                  &   \multirow{2}{*}{CE}                   & -                                                                      & 30                                                                      & 1759                                                                         & 3.3                                                              & 32.5                                                            \\
		&                                                        &                     & -                                                                      & 433                                                                      & 1759                                                                         & 1.4                                                              & 28.6                                                            \\ \cline{2-8} 
		\multirow{2}{*}{AV-HuBERT (w/ self-training)~\cite{shi2022learning}} & \multicolumn{1}{l}{\multirow{2}{*}{Transformer-Large}} & \multirow{2}{*}{CE} & -                                                                      & 30                                                                       & 1759                                                                         & -                                                              & 28.6                                                            \\
		& \multicolumn{1}{l}{}                                   &                     & -                                                                      & 433                                                                      & 1759                                                                         & -                                                              & 26.9                                                            \\ \hline
		\multirow{5}{*}{{\scshape VatLM} (ours)\tnote{*}}                 & \multirow{3}{*}{Transformer-Base}                      & \multirow{3}{*}{CE} & -                                                                      & 30                                                                       & 433                                                                          & \textbf{3.6}                                                             & \textbf{48.0}                                                            \\
		&                                                        &                     & -                                                                      & 30                                                                       & 1759                                                                         & \textbf{3.4}                                                              & \textbf{42.6}                                                            \\
		&                                                        &                     & -                                                                      & 433                                                                      & 1759                                                                         & \textbf{1.7}                                                              & \textbf{34.2}                                                            \\ \cline{2-8} 
		& \multirow{2}{*}{Transformer-Large}                     & \multirow{2}{*}{CE} & -                                                                      & 30                                                                       & 1759                                                                         & \textbf{2.7}                                                              & \textbf{31.6}                                                            \\
		&                                                        &                     & -                                                                      & 433                                                                      & 1759                                                                         & \textbf{1.2}                                                              & \textbf{28.4}                                                            \\ \cline{2-8} 
		\multirow{2}{*}{{\scshape VatLM} (w/ self-training)\tnote{*}}     & \multicolumn{1}{l}{\multirow{2}{*}{Transformer-Large}} & \multirow{2}{*}{CE} & -                                                                      & 30                                                                       & 1759                                                                         & \textbf{2.7}                                                              & \textbf{27.6}                                                            \\
		& \multicolumn{1}{l}{}                                   &                     & -                                                                      & 433                                                                      & 1759                                                                         & \textbf{1.2}                                                              & \textbf{26.2}                                                            \\ \hline
	\end{tabular}

\begin{tablenotes}
\footnotesize
\item[*] In the pre-training stage, our model uses additional 3846h unpaired audio, 452h audio-text and 600M unpaired text data.
\end{tablenotes}

\end{threeparttable}
\end{table*}

\subsection{Pre-Training Setup}
	Our model is implemented using Fairseq~\cite{ott-etal-2019-fairseq}.
	The visual encoder is a modified ResNet-18, the audio encoder uses a linear project layer, and the text encoder contains an embedding layer.
	We consider two model configurations: a 12-layer Transformer block for the base model and a 24-layer Transformer block for the large model.
	For the base and large models, the embedding layer dimension/feed-forward neural network dimension/number of attention heads in each transformer block are 768/3072/12 and 1024/4096/16, respectively.
	The inputs to the visual encoder are lip ROIs, and the inputs to the audio encoder are log-fbank features with 104 dimensions and a frame shift of 40 ms (obtained by concatenating 4 adjacent fbank features with 26 dimensions and a frame shift of 10 ms), and the input to the text encoder is a sequence of phonemes.
	The parameter amounts of the base and large models are 107M and 332M, respectively. It is worth mentioning that the parameters of AV-HuBERT base model and large model are 103M and 325M respectively, so the VATLM model only increases a small amount of parameters compared with AV-HuBERT. It takes about 3 days to train the VATLM Base model with 400k updates using 32 Tesla-V100-32G GPUs and about 6 days to train the VATLM Large model with 600k updates using 32 Tesla-A100-80G GPUs.
	The number of k-means units is set to be 2000.
	Both $\lambda_1$, $\lambda_2$ and $\lambda_3$ in Equa.\ref{eq3} are empirically set to be 1.0. 
	We employ the same modality dropout and input mask strategies as in~\cite{shi2022learning} in the pre-training stage. The model is trained using the Adam optimizer.
	Since we utilize different types of datasets, the mini-batches of our model and AV-HuBERT are kept consistent by controlling the sampling ratio of different data. 
	The phoneme2unit model contains a 4-layer encoder, a 4-layer decoder and a duration predictor module.
	The dimensions of the model are 256. The number of phoneme categories is 41, and the number of unit categories is 2000.
	The phoneme2unit model utilizes the Adam optimizer. We train 40k updates for 433 hours of TED-LIUM3 data, with a learning rate of 5e-4 and a mini-batch of 10K phonemes.

\subsection{Fine-Tuning and Evaluation}
	After pre-training, we fine-tune the {\scshape VatLM} model with labeled data. 
	For the AVSR task, we remove the text encoder and replace its output with zero vectors.
	The base model is followed by a 6-layer Transformer decoder, and the large model is followed by a 9-layer Transformer decoder.
	For the VSR task, we remove the audio encoder and the text encoder and replace their outputs with zero vectors. The decoder is the same as that for the AVSR task.
	The modeling unit for the 30-hour/433-hour AVSR task and the VSR task adopts 1000 subword units, which are obtained using the sentencepiece~\cite{kudosentencepiece} toolkit.
	After fine-tuning, we decode the test set with a beam size of 50 and without any language model.
	The WER is used as the performance metric to evaluate the model.
	To evaluate the performance in noisy scenarios, we utilize 30 hours of noisy labeled data to fine-tune the {\scshape VatLM} base model that is obtained using 433 hours of data for pre-training.
	During fine-tuning, the training data of the LRS3 dataset is dynamically mixed with MUSAN noise data at a signal-to-noise ratio (SNR) in [0,25] dB to generate noisy speech.
	Then we generate the noisy test and validation sets by mixing the noise at SNRs of \{-10, -5, 0, 5, 10\} dB, where the noise types are babble, speech, and music.

\section{Experimental Results}
\label{sec:experiment_results}

\begin{table}[]
	\caption{The performance comparison in WER of various models for AVSR and VSR on the LRS2 dataset.}
	\label{tab:table3}
	\centering
	\begin{tabular}{lccc}
		\hline
		\textbf{Method}    & \textbf{LM} & \textbf{\begin{tabular}[c]{@{}c@{}}VSR\\ WER(\%)\end{tabular}} & \textbf{\begin{tabular}[c]{@{}c@{}}AVSR\\ WER(\%)\end{tabular}} \\ \hline
		\multicolumn{4}{l}{\textbf{Supervised}}                                                                                                                             \\
		LIBS~\cite{zhao2020hearing}               & -           & 65.3                                                           & -                                                               \\
		TM-CTC~\cite{8585066}             & -           & 54.7                                                           & 8.2                                                             \\
		Conv-seq2seq~\cite{Zhang_2019_ICCV}       & -           & 51.7                                                           & -                                                               \\
		TM-seq2seq~\cite{8585066}         & -           & 50.0                                                           & 8.5                                                             \\
		KD-TM~\cite{Ren_2021_CVPR}              & -           & 49.2                                                           & -                                                               \\
		LF-MMI TDNN~\cite{9054127}        & \checkmark   & 48.9                                                           & 5.9                                                             \\
		Kim et al.~\cite{kim2022distinguishing} & - & 44.5 & -
        \\
        Ma et al.~\cite{ma2022visual} & - & 32.9 & -
		\\
		E2E Conformer~\cite{9414567}      & \checkmark   & 37.9                                                           & 3.7                                                             \\ \hline
		\multicolumn{4}{l}{\textbf{Self-supervised}}                                                                                                                        \\
		Pan et al.~\cite{pan-etal-2022-leveraging}         & -           & 43.2                                                           & 2.6                                                             \\
		AV-HuBERT Base~\cite{shi2022learning}     & -           & 31.2                                                           & 3.6                                                             \\
		AV-HuBERT Large~\cite{shi2022learning}    & -           & 25.5                                                           & 2.5                                                             \\
		{\scshape VatLM} Base (ours)  & -           & \textbf{30.6}                                                           & \textbf{2.9}                                                             \\
		{\scshape VatLM} Large (ours) & -           & \textbf{24.3}                                                               &  \textbf{2.3}                                                               \\ \hline
	\end{tabular}
\end{table}

\subsection{Evaluation on AVSR}

\textbf{Comparison methods}: As one of the supervised methods, Afouras et al.~\cite{8585066} employes 1362 hours of labeled data, trained the AVSR model based on the Transformer and sequence-to-sequence cross-entropy (CE)/connectionist temporal classification (CTC) criterion, and decoded the test set using an additional recurrent neural network (RNN) based language model.
Xu et al.~\cite{Xu_2020_CVPR} uses a 3D residual convolution to encode visual information and an element-wise attention gated recurrent unit to model long sequence information.
Ma et al.~\cite{9414567} trains the AVSR model based on the Conformer model under the CTC and CE criterion and decoded the test set using an additional Transformer-based language model.
As a self-supervised approach, 
Zhang et al.~\cite{zhang2022learning} uses a contrastive loss function for self-supervised pre-training.
AV-HuBERT~\cite{shi2022learning} uses both audio and visual features for clustering to generate hidden units, and the masked prediction and fine-tuning are similarly incorporated as HuBERT.

Table~\ref{tab:table1} shows the performance of the proposed {\scshape VatLM} model in comparison to previously published supervised and self-supervised AVSR models on the LRS3 dataset using different amounts of fine-tuning and pre-training data. 
Using 433 hours of unlabeled visual-audio LRS3 data, paired audio data, unpaired text data, and unpaired audio data for pre-training together with 30 hours of transcribed visual-audio data for fine-tuning, our {\scshape VatLM} model can achieve a WER of 3.6 on the AVSR task.
Using 1759 hours of unlabeled visual-audio data, paired audio data, unpaired text data, and unpaired audio data for pre-training together with 30 hours of transcribed visual-audio data for fine-tuning, our {\scshape VatLM} model is capable of achieving a WER of 3.4.
When using 433 hours of transcribed visual-audio data, our {\scshape VatLM} model is able to achieve a WER of 1.7 on the AVSR task, which outperforms existing supervised and self-supervised models.
This shows that pre-training with more labeled/unlabeled unimodal data can improve the performance of downstream multimodal tasks. 
When the large model is used, AVSR performance can be consistently improved and exceeds or is comparable to other methods.

\begin{table}[]
	\caption{Performance comparison of our pre-trained {\scshape VatLM} model using different training losses.}
	\label{tab:table4}
	\centering
	\begin{tabular}{c|clcc}
		\hline
		\# & \textbf{\begin{tabular}[c]{@{}c@{}}Fine-tuned \\ dataset \end{tabular}} & \textbf{\begin{tabular}[c]{@{}c@{}} \ \ \ \ \ Pre-training loss \\ \end{tabular}}        & \textbf{\begin{tabular}[c]{@{}c@{}}VSR\\ WER (\%)\end{tabular}} & \textbf{\begin{tabular}[c]{@{}c@{}}AVSR\\ WER (\%)\end{tabular}} \\ \hline
  	1 & 30 LRS3                                                                  & \begin{tabular}[c]{@{}c@{}} $\mathcal{L}_{total}$ \end{tabular} & 48.0   & 3.6        \\ \hline
		2 & 30 LRS3                                                                   & \begin{tabular}[c]{@{}c@{}} $ \ \ - \mathcal{L}^{p}$ \end{tabular}        & 48.3                                                            & 3.7                                                              \\ \hline
  	3 &	30 LRS3                                                                  & \begin{tabular}[c]{@{}c@{}}$\ \ - \mathcal{L}^{a}$\end{tabular}         & 48.3                                                            & 3.9                                                              \\ \hline
		4 & 30 LRS3                                                                  & $\ \ - \mathcal{L}^{p} - \mathcal{L}^{a}$                                                       & 49.2                                                            & 4.2                                                              \\ \hline 
		5 & 30 LRS3                                                                  & $\ \ - \mathcal{L}^{p} - \mathcal{L}^{a} -  \mathcal{L}^{ap} $                                                                       & 51.8                                                            & 4.9                                                              \\ \hline

		6 & \begin{tabular}[c]{@{}c@{}}30 LRS3 + \\  TED-LIUM3 \end{tabular}         & $\ \ - \mathcal{L}^{p} - \mathcal{L}^{a} -  \mathcal{L}^{ap} $                                                                          & -                                                               & 4.7                                                              \\ \hline

	\end{tabular}
\end{table}

\subsection{Evaluation on VSR}
\textbf{Comparison methods}: In order to show the efficacy of the proposed method on VSR, we also involve several state-of-the-art supervised and self-supervised approaches for comparison. For instance, 
Zhang et al.~\cite{Zhang_2019_ICCV} utilizes the spatial-temporal fusion module to model the short-range temporal dependence and local spatial information of lip sequences, which depends on extra 157 hours of labeled LRW data to train the model.
Afouras et al.~\cite{8585066} uses 1362 hours of labeled data to train a supervised VSR model.
Shillingford et al.~\cite{shillingford19_interspeech} constructs 3889 hours of video clips of transcribed speech faces and trained a VSR model using an RNN-based model structure and CTC loss function.
Makino et al.~\cite{9004036} constructs 31k hours of transcribed audio-visual data by extracting videos from YouTube and trained an RNN-based VSR model.
For semi-supervised and self-supervised methods, we mainly choose the models from~\cite{9054253,ma21c_interspeech,shi2022learning,zhang2022learning}. 

Table~\ref{tab:table1} lists the performance of our {\scshape VatLM} model and previously published supervised and self-supervised VSR models on the LRS3 dataset using different amounts of fine-tuning and pre-training data. 
Using 433 hours of unlabeled visual-audio data, paired audio data, unpaired text data, and unpaired audio data for pre-training as well as 30 hours of transcribed visual data for fine-tuning, our {\scshape VatLM} model is capable of achieving a WER of 48.0.
Using more unlabeled visual-audio data, the VSR performance of {\scshape VatLM} can be improved, which implies that pre-training with more visual-audio data is beneficial for VSR, particularly under low-resource conditions.
Comparing the WER (46.1 vs. 42.6) performance of the VSR task between AV-HuBERT and {\scshape VatLM}, the {\scshape VatLM} model does not utilize more visual data and has further performance gains on VSR, indicating that the use of visual irrelevant data in the pre-training stage can also improve the performance of the VSR task.
We believe that the performance of modality-independent tasks can be improved to some extent when the data of different modalities can be aligned to the same space (see~\ref{sec:VisualizationAnalysis}).
When using 433 hours of transcribed visual data for fine-tuning, our {\scshape VatLM} model can achieve a WER of 34.2 on the VSR task, which becomes comparable to AV-HuBERT. 
The performance of VSR can be consistently improved by using large model compared to using base model.
When the self-training approach is used to decode the untranscribed Voxceleb2 dataset to obtain pseudo labels and add them to the training set, our model achieves a WER of 26.2 on the 433-hour VSR task.
It is worth mentioning that short sentences are not beneficial for the VSR task. Filtering out short sentences smaller than two seconds can improve the quality of pseudo labels.

\begin{table*}[]
	\caption{ The performance comparison in WER between our model and AV-HuBERT model under different SNR noise conditions.}
	\label{tab:table5}
	\centering
	\resizebox{0.95\textwidth}{!}{
		\begin{tabular}{l|c|cccccc|cccccc|cccccc|c}
			\hline
			\multirow{2}{*}{\textbf{Model}}     & \multirow{2}{*}{\textbf{Mode}} & \multicolumn{6}{c|}{\textbf{Babble, SNR=}}       & \multicolumn{6}{c|}{\textbf{Speech, SNR=}}        & \multicolumn{6}{c|}{\textbf{Music, SNR=}}       & \textbf{Clean} \\
			&                       & -10   & -5   & 0    & 5    & 10  & avg  & -10   & -5   & 0    & 5    & 10   & avg  & -10  & -5   & 0    & 5    & 10  & avg  & $\infty$      \\ \hline
			\multirow{2}{*}{AV-HuBERT} & ASR                     & 120.1 & 93.2 & 44.7 & 18.1 & 9.6 & 57.1 & 120.0 & 93.0 & 59.2 & 30.9 & 15.0 & 63.6 & 72.1 & 44.9 & 21.8 & 12.3 & 8.4 & 31.9 & 6.2   \\
			& AVSR                    & 49.9  & 34.4 & 18.8 & 10.4 & 7.7 & 24.2 & 49.7  & 34.5 & 16.9 & 11.5 & 7.9  & 24.1 & 29.9 & 19.2 & 11.9 & 8.8  & 6.7 & 15.3 & 4.9   \\ \hline
			\multirow{2}{*}{{\scshape VatLM}}     & ASR                     & 118.0 & 91.4 & 34.1 & 12.3 & 6.6 & 52.5 & 117.9 & 91.5 & 41.7 & 20.4 & 10.3 & 56.4 & 65.0 & 35.2 & 14.7 & 7.9  & 5.7 & 25.7 & 4.1   \\
			& AVSR                    & 45.8  & 30.6 & 13.4 & 6.8  & 4.9 & 20.3 & 45.6  & 30.8 & 10.9 & 7.1  & 5.0  & 19.9 & 25.2 & 13.9 & 7.7  & 5.5  & 4.8 & 11.4 & 3.6   \\ \hline
	\end{tabular}}
\end{table*}

\begin{figure*}[]
	\centering
	\includegraphics[width=0.9\textwidth]{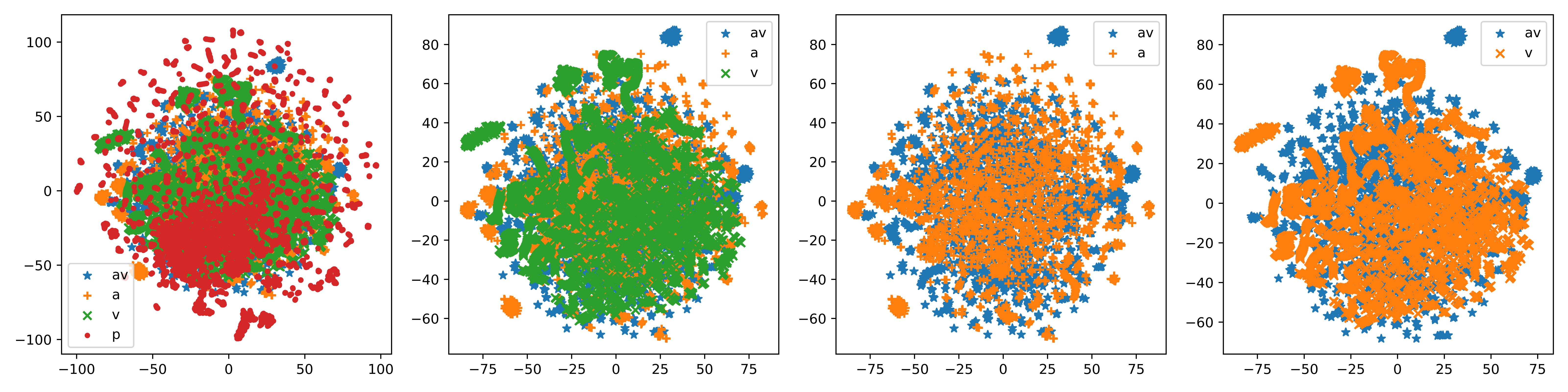}
	\caption{The 2D t-SNE visualization of representations obtained from data with different modalities, where `av' in the figure denotes data of audio-visual modality, `a' denotes data of audio modality, `v' denotes data of visual modality, and `p' denotes data of text modality }
	\label{fig:tsne}
\end{figure*}

\subsection{More Evaluation on LRS2 Dataset}

To verify the generality of the pre-trained model, we fine-tune the {\scshape VatLM} model on the transcribed LRS2 dataset, which is not used in the pre-training phase.
The corresponding results are shown in Table~\ref{tab:table3}.
We can see that the supervised method E2E conformer~\cite{9414567} shows excellent performance, e.g., a WER of 3.7 on AVSR and a WER of 37.9 on VSR, however it incorporates an additional language model for decoding.
For the self-supervised approach, Pan et al.~\cite{pan-etal-2022-leveraging} uses a unimodal visual pre-trained model and a unimodal audio pre-trained model to improve the AVSR/VSR performance.
Since the results of AV-HuBERT were not reported on the LRS2 dataset, we fine-tuned the AV-HuBERT Base model, which reaches a WER of 3.6 on AVSR and a WER of 31.2 on VSR.
The performance of the AV-HuBERT model on the VSR task is significantly improved compared to previous work.
Our {\scshape VatLM} Base model achieves a WER of 2.9 on the AVSR task and 30.6 on the VSR task, with a large performance gain for the AVSR task and a small gain for the VSR task compared to the AV-HuBERT Base model. 
When using the large model, the AVSR and VSR tasks are able to achieve WERs of 24.3 and 2.3, respectively, which is the best-known performance. 
Experimental results show that utilizing multimodal data in the pre-training phase can improve the generalizability of the model for downstream tasks.

\subsection{Ablation Study}
To better illustrate the impact of using different training objectives (as Equa. \ref{eq3}) in the pre-training stage on the performance of downstream tasks, we conduct comparative experiments and the results are shown in Table~\ref{tab:table4}.
We list the performance of our {\scshape VatLM} in line 1 with pre-training loss $\mathcal{L}_{total}$, including $\mathcal{L}^{av}$, $\mathcal{L}^a$, $\mathcal{L}^{ap}$, and $\mathcal{L}^p$.
On this basis, we first individually remove the loss $\mathcal{L}^p$ or $\mathcal{L}^a$ from $\mathcal{L}_{total}$. The results reported in line 2 and line 3 demonstrate that without pre-training of unpaired audio or text data, the performance of AVSR and VSR degrades, indicating the usefulness of $\mathcal{L}^p$ and $\mathcal{L}^a$.
Besides, the combination of $\mathcal{L}^p$ and $\mathcal{L}^a$ is more beneficial for {\scshape VatLM}, because WER rises significantly when removing them simultaneously.
We further pre-train the model without $\mathcal{L}^{ap}$ (line 5), which achieves 51.8\% WER and 4.9\% WER, and underperforms the quality of line 4, demonstrating the importance of $\mathcal{L}^{ap}$.
Finally, we utilize additional paired audio-text data (TED-LIUM3) together with 30 hours of transcribed visual-audio data (LRS3) to fine-tune the pre-trained model of line 5, resulting in a WER of 4.7 on the AVSR task.
Comparing the results of line 6 and line 4, which utilizes the paired audio-text data TED-LIUM3 in the pre-training stage, we can draw a conclusion that 
in case additional paired audio-text data are available, performing a uniform masked prediction task in the pre-training phase makes fuller usage of the data than fine-tuning in the downstream task.


\subsection{Robustness Evaluation}
To evaluate the noise robustness of the model in different noise scenarios and the effect of different modalities on the model performance, we utilize 30 hours of noisy data to fine-tune the pre-trained model obtained from 433 hours of data. Note that we do not employ noise augmentation in the pre-training stage.
We then test on the babble, speech, and music noisy datasets with \{-10, -5, 0, 5, 10\} dB, and the experimental results are shown in Table~\ref{tab:table5}.
As can be seen from the table, our {\scshape VatLM} model outperforms the AV-HuBERT model under the same noise type and SNRs. 
In addition, the model with visual-audio modality (AVSR) has a lower WER than the model with a single audio modality (ASR) under the noise scenario, which indicates that the multimodal model has an advantage compared to the single-modal model in the noise scenario.

\subsection{Visualization Analysis}
\label{sec:VisualizationAnalysis}
Our {\scshape VatLM} model attempts to map data from different modalities to the same space. To analyze whether the data of different modalities are aligned in one space, we sample the data of different modalities from the test set. Then the representations corresponding to different modalities output from the Transformer encoder are reduced to 2D using t-SNE for visualization, as shown in Fig.~\ref{fig:tsne}.
The distribution of audio-visual, audio, visual, and text representations in Fig.~\ref{fig:tsne} is approximately the same, indicating that the representations of different modalities are aligned in the same space. 
When data from different modalities can be aligned in one space, it is possible to improve the performance of tasks with different modalities using data from a single modality.



\section{Conclusion}
\label{sec:conclusion}

In this paper, we present {\scshape VatLM}, a visual-audio-text self-supervised speech representation learning model, which achieves competitive even state-of-the-art performance on audio-visual/visual speech recognition tasks.
To the best of our knowledge, this is the first work to integrate both visual information and textual information into speech representation learning in a unified framework. We tokenize the data sources from different modalities into shared hidden units, and employ a multipath Transformer to model the latent representation of visual, speech, and text. Particularly, the model is optimized by learning to predict the hidden units of different modalities with a unified masked prediction task.
For future work, we are planning to extend to more downstream tasks and model a rich variety of videos with the proposed {\scshape VatLM}. We are also interested in integrating masked autoencoder methods into our model.

\section{discussion}
\label{sec:discussion}

Although the proposed VATLM model can improve speech representation using resources from different modalities (e.g., visual-audio pairs, audio-text pairs, unlabeled speech and unlabeled text), it still has some limitations: (1) The phoneme2unit model requires a small amount of paired ASR data for training. (2) For the AVSR and VSR tasks, using additional unlabeled speech data is more effective than using additional text data, and we conjecture that the text modality is more disparate from the visual modality. VSR tasks are still challenging at the current time, with issues such as homophones. 
(3) How to better utilize data from text modality in visual-speech-text modality representation learning needs further investigation. (4) Methods that use discrete units as self-supervised pre-training targets (e.g., HuBERT, AV-HuBERT and VATLM) require several iterations, so we need to explore more time-efficient training methods.

\bibliographystyle{IEEEtran}
\bibliography{strings,refs}

\begin{thebibliography}{10}
\providecommand{\url}[1]{#1}
\csname url@samestyle\endcsname
\providecommand{\newblock}{\relax}
\providecommand{\bibinfo}[2]{#2}
\providecommand{\BIBentrySTDinterwordspacing}{\spaceskip=0pt\relax}
\providecommand{\BIBentryALTinterwordstretchfactor}{4}
\providecommand{\BIBentryALTinterwordspacing}{\spaceskip=\fontdimen2\font plus
\BIBentryALTinterwordstretchfactor\fontdimen3\font minus
  \fontdimen4\font\relax}
\providecommand{\BIBforeignlanguage}[2]{{%
\expandafter\ifx\csname l@#1\endcsname\relax
\typeout{** WARNING: IEEEtran.bst: No hyphenation pattern has been}%
\typeout{** loaded for the language `#1'. Using the pattern for}%
\typeout{** the default language instead.}%
\else
\language=\csname l@#1\endcsname
\fi
#2}}
\providecommand{\BIBdecl}{\relax}
\BIBdecl

\bibitem{865479}
S.~Dupont and J.~Luettin, ``Audio-visual speech modeling for continuous speech
  recognition,'' \emph{IEEE Trans. Multimedia}, vol.~2, no.~3, pp. 141--151,
  2000.

\bibitem{6384798}
H.~Li, Y.~Wang, T.~Mei, J.~Wang, and S.~Li, ``Interactive multimodal visual
  search on mobile device,'' \emph{IEEE Trans. Multimedia}, vol.~15, no.~3, pp.
  594--607, 2013.

\bibitem{zhu2021deep}
H.~Zhu, M.-D. Luo, R.~Wang, A.-H. Zheng, and R.~He, ``Deep audio-visual
  learning: A survey,'' \emph{Int. J. Autom. Comput.}, vol.~18, no.~3, pp.
  351--376, 2021.

\bibitem{9566778}
M.~Kim, J.~Hong, S.~J. Park, and Y.~M. Ro, ``Cromm-vsr: Cross-modal memory
  augmented visual speech recognition,'' \emph{IEEE Trans. Multimedia},
  vol.~24, pp. 4342--4355, 2022.

\bibitem{mohamed2022self}
A.~Mohamed, H.-y. Lee, L.~Borgholt, J.~D. Havtorn, J.~Edin, C.~Igel,
  K.~Kirchhoff, S.-W. Li, K.~Livescu, L.~Maal{\o}e \emph{et~al.},
  ``Self-supervised speech representation learning: A review,'' \emph{arXiv
  preprint arXiv:2205.10643}, 2022.

\bibitem{Schneider2019}
S.~Schneider, A.~Baevski, R.~Collobert, and M.~Auli, ``{Wav2vec: Unsupervised
  Pre-Training for Speech Recognition},'' in \emph{ISCA Interspeech}, 2019, pp.
  3465--3469.

\bibitem{NEURIPS2020_92d1e1eb}
A.~Baevski, Y.~Zhou, A.~Mohamed, and M.~Auli, ``Wav2vec 2.0: A framework for
  self-supervised learning of speech representations,'' in \emph{Proc. Adv.
  Neural Inf. Process. Syst.}, 2020, pp. 12\,449--12\,460.

\bibitem{9585401}
W.-N. Hsu, B.~Bolte, Y.-H.~H. Tsai, K.~Lakhotia, R.~Salakhutdinov, and
  A.~Mohamed, ``Hubert: Self-supervised speech representation learning by
  masked prediction of hidden units,'' \emph{IEEE/ACM Trans. Audio, Speech,
  Lang. Process.}, vol.~29, pp. 3451--3460, 2021.

\bibitem{9814838}
S.~Chen, C.~Wang, Z.~Chen, Y.~Wu, S.~Liu, Z.~Chen, J.~Li, N.~Kanda,
  T.~Yoshioka, X.~Xiao, J.~Wu, L.~Zhou, S.~Ren, Y.~Qian, Y.~Qian, J.~Wu,
  M.~Zeng, X.~Yu, and F.~Wei, ``Wavlm: Large-scale self-supervised pre-training
  for full stack speech processing,'' \emph{IEEE J. of Selected Topics in
  Signal Process.}, pp. 1--14, 2022.

\bibitem{baevski2022data2vec}
A.~Baevski, W.~Hsu, Q.~Xu, A.~Babu, J.~Gu, and M.~Auli, ``data2vec: A general
  framework for self-supervised learning in speech, vision and language,'' in
  \emph{Int. Conf. Machine Learning (ICML)}, vol. 162, 2022, pp. 1298--1312.

\bibitem{oord2018representation}
A.~v.~d. Oord, Y.~Li, and O.~Vinyals, ``Representation learning with
  contrastive predictive coding,'' \emph{arXiv preprint arXiv:1807.03748},
  2018.

\bibitem{liu2021tera}
A.~T. Liu, S.-W. Li, and H.-y. Lee, ``Tera: Self-supervised learning of
  transformer encoder representation for speech,'' \emph{IEEE/ACM Trans. Audio,
  Speech, Lang. Process.}, vol.~29, pp. 2351--2366, 2021.

\bibitem{Chung2019}
Y.-A. Chung, W.-N. Hsu, H.~Tang, and J.~Glass, ``{An Unsupervised
  Autoregressive Model for Speech Representation Learning},'' in \emph{ISCA
  Interspeech}, 2019, pp. 146--150.

\bibitem{xu2022masked}
H.~Xu, J.~Li, A.~Baevski, M.~Auli, W.~Galuba, F.~Metze, C.~Feichtenhofer
  \emph{et~al.}, ``Masked autoencoders that listen,'' \emph{arXiv preprint
  arXiv:2207.06405}, 2022.

\bibitem{vaswani2017attention}
A.~Vaswani, N.~Shazeer, N.~Parmar, J.~Uszkoreit, L.~Jones, A.~N. Gomez,
  {\L}.~Kaiser, and I.~Polosukhin, ``Attention is all you need,'' \emph{Proc.
  Adv. Neural Inf. Process. Syst.}, vol.~30, pp. 6000--6010, 2017.

\bibitem{9688253}
Y.-A. Chung, Y.~Zhang, W.~Han, C.-C. Chiu, J.~Qin, R.~Pang, and Y.~Wu,
  ``w2v-bert: Combining contrastive learning and masked language modeling for
  self-supervised speech pre-training,'' in \emph{IEEE Autom. Speech Recognit.
  Understanding Workshop}, 2021, pp. 244--250.

\bibitem{he2022masked}
K.~He, X.~Chen, S.~Xie, Y.~Li, P.~Doll{\'a}r, and R.~Girshick, ``Masked
  autoencoders are scalable vision learners,'' in \emph{Proc. IEEE/CVF Conf.
  Comput. Vis. Pattern Recognit.}, 2022, pp. 16\,000--16\,009.

\bibitem{Xu_2020_CVPR}
B.~Xu, C.~Lu, Y.~Guo, and J.~Wang, ``Discriminative multi-modality speech
  recognition,'' in \emph{Proc. IEEE/CVF Conf. Comput. Vis. Pattern Recognit.},
  2020, pp. 14\,421--14\,430.

\bibitem{9004036}
T.~Makino, H.~Liao, Y.~Assael, B.~Shillingford, B.~Garcia, O.~Braga, and
  O.~Siohan, ``Recurrent neural network transducer for audio-visual speech
  recognition,'' in \emph{IEEE Autom. Speech Recognit. Understanding Workshop},
  2019, pp. 905--912.

\bibitem{bapna2022mslam}
A.~Bapna, C.~Cherry, Y.~Zhang, Y.~Jia, M.~Johnson, Y.~Cheng, S.~Khanuja,
  J.~Riesa, and A.~Conneau, ``mslam: Massively multilingual joint pre-training
  for speech and text,'' \emph{arXiv preprint arXiv:2202.01374}, 2022.

\bibitem{zhang2022speechlm}
Z.~Zhang, S.~Chen, L.~Zhou, Y.~Wu, S.~Ren, S.~Liu, Z.~Yao, X.~Gong, L.~Dai,
  J.~Li \emph{et~al.}, ``Speechlm: Enhanced speech pre-training with unpaired
  textual data,'' \emph{arXiv preprint arXiv:2209.15329}, 2022.

\bibitem{sumby1954visual}
W.~H. Sumby and I.~Pollack, ``Visual contribution to speech intelligibility in
  noise,'' \emph{J. Acoust. Soc. Amer.}, vol.~26, no.~2, pp. 212--215, 1954.

\bibitem{8461326}
S.~Petridis, T.~Stafylakis, P.~Ma, F.~Cai, G.~Tzimiropoulos, and M.~Pantic,
  ``End-to-end audiovisual speech recognition,'' in \emph{IEEE Int. Conf.
  Acoust., Speech, Signal Process.}, 2018, pp. 6548--6552.

\bibitem{8585066}
T.~Afouras, J.~S. Chung, A.~Senior, O.~Vinyals, and A.~Zisserman, ``Deep
  audio-visual speech recognition,'' \emph{IEEE Trans. Pattern Anal. Mach.
  Intell.}, pp. 1--1, 2018.

\bibitem{9414567}
P.~Ma, S.~Petridis, and M.~Pantic, ``End-to-end audio-visual speech recognition
  with conformers,'' in \emph{IEEE Int. Conf. Acoust., Speech, Signal
  Process.}, 2021, pp. 7613--7617.

\bibitem{9018185}
F.~Tao and C.~Busso, ``End-to-end audiovisual speech recognition system with
  multitask learning,'' \emph{IEEE Trans. Multimedia}, vol.~23, pp. 1--11,
  2021.

\bibitem{4540195}
J.-S. Lee and C.~H. Park, ``Robust audio-visual speech recognition based on
  late integration,'' \emph{IEEE Trans. Multimedia}, vol.~10, no.~5, pp.
  767--779, 2008.

\bibitem{bapna2021slam}
A.~Bapna, Y.-a. Chung, N.~Wu, A.~Gulati, Y.~Jia, J.~H. Clark, M.~Johnson,
  J.~Riesa, A.~Conneau, and Y.~Zhang, ``Slam: A unified encoder for speech and
  language modeling via speech-text joint pre-training,'' \emph{arXiv preprint
  arXiv:2110.10329}, 2021.

\bibitem{ao-etal-2022-speecht5}
J.~Ao, R.~Wang, L.~Zhou, C.~Wang, S.~Ren, Y.~Wu, S.~Liu, T.~Ko, Q.~Li,
  Y.~Zhang, Z.~Wei, Y.~Qian, J.~Li, and F.~Wei, ``{S}peech{T}5: Unified-modal
  encoder-decoder pre-training for spoken language processing,'' in \emph{Proc.
  Annu. Meeting Assoc. Comput. Linguistics}, 2022, pp. 5723--5738.

\bibitem{shi2022learning}
B.~Shi, W.-N. Hsu, K.~Lakhotia, and A.~Mohamed, ``Learning audio-visual speech
  representation by masked multimodal cluster prediction,'' \emph{arXiv
  preprint arXiv:2201.02184}, 2022.

\bibitem{hsu2022single}
W.-N. Hsu and B.~Shi, ``A single self-supervised model for many speech
  modalities enables zero-shot modality transfer,'' \emph{arXiv preprint
  arXiv:2207.07036}, 2022.

\bibitem{pan-etal-2022-leveraging}
X.~Pan, P.~Chen, Y.~Gong, H.~Zhou, X.~Wang, and Z.~Lin, ``Leveraging unimodal
  self-supervised learning for multimodal audio-visual speech recognition,'' in
  \emph{Proc. Annu. Meeting Assoc. Comput. Linguistics}, 2022, pp. 4491--4503.

\bibitem{zhang2022learning}
Z.-Q. Zhang, J.~Zhang, J.-S. Zhang, M.-H. Wu, X.~Fang, and L.-R. Dai,
  ``Learning contextually fused audio-visual representations for audio-visual
  speech recognition,'' in \emph{IEEE Int. Conf. Image Process.}, 2022, pp.
  1346--1350.

\bibitem{shi2022robust}
B.~Shi, W.-N. Hsu, and A.~Mohamed, ``{Robust Self-Supervised Audio-Visual
  Speech Recognition},'' in \emph{ISCA Interspeech}, 2022, pp. 2118--2122.

\bibitem{tang-etal-2022-unified}
Y.~Tang, H.~Gong, N.~Dong, C.~Wang, W.-N. Hsu, J.~Gu, A.~Baevski, X.~Li,
  A.~Mohamed, M.~Auli, and J.~Pino, ``Unified speech-text pre-training for
  speech translation and recognition,'' in \emph{Proc. Annu. Meeting Assoc.
  Comput. Linguistics}, 2022, pp. 1488--1499.

\bibitem{weiwang2021}
W.~Wang, S.~Ren, Y.~Qian, S.~Liu, Y.~Shi, Y.~Qian, and M.~Zeng, ``Optimizing
  alignment of speech and language latent spaces for end-to-end speech
  recognition and understanding,'' in \emph{IEEE Int. Conf. Acoust., Speech,
  Signal Process.}, 2022, pp. 7802--7806.

\bibitem{wang2022image}
W.~Wang, H.~Bao, L.~Dong, J.~Bjorck, Z.~Peng, Q.~Liu, K.~Aggarwal, O.~K.
  Mohammed, S.~Singhal, S.~Som \emph{et~al.}, ``Image as a foreign language:
  Beit pretraining for all vision and vision-language tasks,'' \emph{arXiv
  preprint arXiv:2208.10442}, 2022.

\bibitem{liu2021opt}
J.~Liu, X.~Zhu, F.~Liu, L.~Guo, Z.~Zhao, M.~Sun, W.~Wang, H.~Lu, S.~Zhou,
  J.~Zhang \emph{et~al.}, ``Opt: omni-perception pre-trainer for cross-modal
  understanding and generation,'' \emph{arXiv preprint arXiv:2107.00249}, 2021.

\bibitem{yang2022code}
Z.~Yang, Y.~Fang, C.~Zhu, R.~Pryzant, D.~Chen, Y.~Shi, Y.~Xu, Y.~Qian, M.~Gao,
  Y.-L. Chen \emph{et~al.}, ``i-code: An integrative and composable multimodal
  learning framework,'' \emph{arXiv preprint arXiv:2205.01818}, 2022.

\bibitem{yang21c_interspeech}
S.~wen Yang, P.-H. Chi, Y.-S. Chuang, C.-I.~J. Lai, K.~Lakhotia, Y.~Y. Lin,
  A.~T. Liu, J.~Shi, X.~Chang, G.-T. Lin, T.-H. Huang, W.-C. Tseng, K.~tik Lee,
  D.-R. Liu, Z.~Huang, S.~Dong, S.-W. Li, S.~Watanabe, A.~Mohamed, and
  H.~yi~Lee, ``{SUPERB: Speech Processing Universal PERformance Benchmark},''
  in \emph{ISCA Interspeech}, 2021, pp. 1194--1198.

\bibitem{9054253}
T.~Afouras, J.~S. Chung, and A.~Zisserman, ``Asr is all you need: Cross-modal
  distillation for lip reading,'' in \emph{IEEE Int. Conf. Acoust., Speech,
  Signal Process.}, 2020, pp. 2143--2147.

\bibitem{ma21c_interspeech}
P.~Ma, R.~Mira, S.~Petridis, B.~W. Schuller, and M.~Pantic, ``{LiRA: Learning
  Visual Speech Representations from Audio Through Self-Supervision},'' in
  \emph{ISCA Interspeech}, 2021, pp. 3011--3015.

\bibitem{9746449}
D.~M. Chan, S.~Ghosh, D.~Chakrabarty, and B.~Hoffmeister, ``Multi-modal
  pre-training for automated speech recognition,'' in \emph{IEEE Int. Conf.
  Acoust., Speech, Signal Process.}, 2022, pp. 246--250.

\bibitem{Prajwal_2022_CVPR}
K.~R. Prajwal, T.~Afouras, and A.~Zisserman, ``Sub-word level lip reading with
  visual attention,'' in \emph{Proceedings of the IEEE/CVF Conference on
  Computer Vision and Pattern Recognition (CVPR)}, June 2022, pp. 5162--5172.

\bibitem{kim2022distinguishing}
M.~Kim, J.~H. Yeo, and Y.~M. Ro, ``Distinguishing homophenes using multi-head
  visual-audio memory for lip reading,'' in \emph{Proceedings of the AAAI
  Conference on Artificial Intelligence}, vol.~36, no.~1, 2022, pp. 1174--1182.

\bibitem{ma2022visual}
P.~Ma, S.~Petridis, and M.~Pantic, ``Visual speech recognition for multiple
  languages in the wild,'' \emph{Nature Machine Intelligence}, pp. 1--10, 2022.

\bibitem{chung-etal-2021-splat}
Y.-A. Chung, C.~Zhu, and M.~Zeng, ``{SPLAT}: Speech-language joint pre-training
  for spoken language understanding,'' in \emph{Proc. Annu. Meeting Assoc.
  Comput. Linguistics}, 2021, pp. 1897--1907.

\bibitem{zhang2020transformer}
Q.~Zhang, H.~Lu, H.~Sak, A.~Tripathi, E.~McDermott, S.~Koo, and S.~Kumar,
  ``Transformer transducer: A streamable speech recognition model with
  transformer encoders and rnn-t loss,'' in \emph{IEEE Int. Conf. Acoust.,
  Speech, Signal Process.}, 2020, pp. 7829--7833.

\bibitem{chen2022maestro}
Z.~Chen, Y.~Zhang, A.~Rosenberg, B.~Ramabhadran, P.~J. Moreno, A.~Bapna, and
  H.~Zen, ``{MAESTRO: Matched Speech Text Representations through Modality
  Matching},'' in \emph{ISCA Interspeech}, 2022, pp. 4093--4097.

\bibitem{akbari2021vatt}
H.~Akbari, L.~Yuan, R.~Qian, W.-H. Chuang, S.-F. Chang, Y.~Cui, and B.~Gong,
  ``Vatt: Transformers for multimodal self-supervised learning from raw video,
  audio and text,'' \emph{Proc. Adv. Neural Inf. Process. Syst.}, vol.~34, pp.
  24\,206--24\,221, 2021.

\bibitem{king2009dlib}
D.~E. King, ``Dlib-ml: A machine learning toolkit,'' \emph{J. Mach. Learn.
  Res.}, vol.~10, pp. 1755--1758, 2009.

\bibitem{martinez2020lipreading}
B.~Martinez, P.~Ma, S.~Petridis, and M.~Pantic, ``Lipreading using temporal
  convolutional networks,'' in \emph{IEEE Int. Conf. Acoust., Speech, Signal
  Process.}, 2020, pp. 6319--6323.

\bibitem{zhang2022speechut}
Z.~Zhang, L.~Zhou, J.~Ao, S.~Liu, L.~Dai, J.~Li, and F.~Wei, ``Speechut:
  Bridging speech and text with hidden-unit for encoder-decoder based
  speech-text pre-training,'' \emph{arXiv preprint arXiv:2210.03730}, 2022.

\bibitem{NEURIPS2019_f63f65b5}
Y.~Ren, Y.~Ruan, X.~Tan, T.~Qin, S.~Zhao, Z.~Zhao, and T.-Y. Liu, ``Fastspeech:
  Fast, robust and controllable text to speech,'' in \emph{Proc. Adv. Neural
  Inf. Process. Syst.}, vol.~32, 2019.

\bibitem{afouras2018lrs3}
T.~Afouras, J.~S. Chung, and A.~Zisserman, ``Lrs3-ted: a large-scale dataset
  for visual speech recognition,'' \emph{arXiv preprint arXiv:1809.00496},
  2018.

\bibitem{chung2018voxceleb2}
J.~S. Chung, A.~Nagrani, and A.~Zisserman, ``Voxceleb2: Deep speaker
  recognition,'' in \emph{ISCA Interspeech}, 2018, pp. 1086--1090.

\bibitem{chen2021gigaspeech}
G.~Chen, S.~Chai, G.-B. Wang, J.~Du, W.-Q. Zhang, C.~Weng, D.~Su, D.~Povey,
  J.~Trmal, J.~Zhang, M.~Jin, S.~Khudanpur, S.~Watanabe, S.~Zhao, W.~Zou,
  X.~Li, X.~Yao, Y.~Wang, Z.~You, and Z.~Yan, ``{GigaSpeech: An Evolving,
  Multi-Domain ASR Corpus with 10,000 Hours of Transcribed Audio},'' in
  \emph{ISCA Interspeech}, 2021, pp. 3670--3674.

\bibitem{tedlium3}
F.~Hernandez, V.~Nguyen, S.~Ghannay, N.~Tomashenko, and Y.~Est{\`e}ve,
  ``Ted-lium 3: Twice as much data and corpus repartition for experiments on
  speaker adaptation,'' in \emph{Speech and Computer}, 2018, pp. 198--208.

\bibitem{7179001}
W.~Williams, N.~Prasad, D.~Mrva, T.~Ash, and T.~Robinson, ``Scaling recurrent
  neural network language models,'' in \emph{IEEE Int. Conf. Acoust., Speech,
  Signal Process.}, 2015, pp. 5391--5395.

\bibitem{shi22_interspeech}
B.~Shi, W.-N. Hsu, and A.~Mohamed, ``{Robust Self-Supervised Audio-Visual
  Speech Recognition},'' in \emph{ISCA Interspeech}, 2022, pp. 2118--2122.

\bibitem{snyder2015musan}
D.~Snyder, G.~Chen, and D.~Povey, ``Musan: A music, speech, and noise corpus,''
  \emph{arXiv preprint arXiv:1510.08484}, 2015.

\bibitem{Zhang_2019_ICCV}
X.~Zhang, F.~Cheng, and S.~Wang, ``Spatio-temporal fusion based convolutional
  sequence learning for lip reading,'' in \emph{Proc. IEEE/CVF Int. Conf.
  Comput. Vis.}, 2019, pp. 713--722.

\bibitem{shillingford19_interspeech}
B.~Shillingford, Y.~Assael, M.~W. Hoffman, T.~Paine, C.~Hughes, U.~Prabhu,
  H.~Liao, H.~Sak, K.~Rao, L.~Bennett, M.~Mulville, M.~Denil, B.~Coppin,
  B.~Laurie, A.~Senior, and N.~de~Freitas, ``{Large-Scale Visual Speech
  Recognition},'' in \emph{ISCA Interspeech}, 2019, pp. 4135--4139.

\bibitem{ott-etal-2019-fairseq}
M.~Ott, S.~Edunov, A.~Baevski, A.~Fan, S.~Gross, N.~Ng, D.~Grangier, and
  M.~Auli, ``fairseq: A fast, extensible toolkit for sequence modeling,'' in
  \emph{Proc. Annu. Meeting Assoc. Comput. Linguistics}, 2019, pp. 48--53.

\bibitem{kudosentencepiece}
T.~Kudo and J.~Richardson, ``{S}entence{P}iece: A simple and language
  independent subword tokenizer and detokenizer for neural text processing,''
  in \emph{Proceedings of the 2018 Conference on Empirical Methods in Natural
  Language Processing: System Demonstrations}, Nov. 2018, pp. 66--71.

\bibitem{zhao2020hearing}
Y.~Zhao, R.~Xu, X.~Wang, P.~Hou, H.~Tang, and M.~Song, ``Hearing lips:
  Improving lip reading by distilling speech recognizers,'' in \emph{Proc. AAAI
  Conf. Artif. Intell.}, vol.~34, no.~04, 2020, pp. 6917--6924.

\bibitem{Ren_2021_CVPR}
S.~Ren, Y.~Du, J.~Lv, G.~Han, and S.~He, ``Learning from the master: Distilling
  cross-modal advanced knowledge for lip reading,'' in \emph{Proc. IEEE/CVF
  Conf. Comput. Vis. Pattern Recognit.}, 2021, pp. 13\,325--13\,333.

\bibitem{9054127}
J.~Yu, S.-X. Zhang, J.~Wu, S.~Ghorbani, B.~Wu, S.~Kang, S.~Liu, X.~Liu,
  H.~Meng, and D.~Yu, ``Audio-visual recognition of overlapped speech for the
  lrs2 dataset,'' in \emph{IEEE Int. Conf. Acoust., Speech, Signal Process.},
  2020, pp. 6984--6988.

\end{thebibliography}
\end{document}